\documentclass[prd,preprint,tightenlines,floatfix,preprintnumbers,nofootinbib,eqsecnum]{revtex4}
 \usepackage[dvips,final]{graphicx}
  \usepackage{amssymb}
   \usepackage{amsmath}
    \usepackage{amsfonts}
     \usepackage{epsfig}
      \usepackage{bm}

\def\QED{\text{{\rm QED}}}
\def\Va{\text{{\rm V}}}
\def\Ha{\text{{\rm H}}}
\def\Da{\text{{\rm D}}}
\def\Ea{\text{{\rm E}}}
\def\Ma{\text{{\rm M}}}
\def\sa{\text{{\rm s}}}
\def\pt{\text{{\rm pt}}}
\def\PT{\text{{\rm PT}}}

\def\APT{\text{{\rm APT}}}
\def\had{\text{{\rm had}}}
\def\expt{\text{{\rm expt}}}
\def\rad{\text{{\rm rad}}}
\def\Ima{\operatorname{Im}}
\def\Req{\operatorname{Re}}

\begin{document}
\thispagestyle{empty} \preprint{\hbox{}} \vspace*{-10mm}

\title{Ten years of the Analytic Perturbation Theory in QCD \footnote{To be
published in Theor. Math. Phys. (2007) in the issue dedicated to
80th birthday of Anatolij Alekseevich Logunov.}}

\author{D.~V.~Shirkov}

\author{I.~L.~Solovtsov}
\altaffiliation[~Also at ]{International Center for Advanced
Studies, Gomel State Technical University, Gomel, 246746, Belarus}

\affiliation{Bogoliubov Laboratory of Theoretical Physics, Joint
Institute for Nuclear Research, Dubna, 141980, Russia}

\begin{abstract}
The renormalization group method enables one to improve the
properties of the QCD perturbative power series in the ultraviolet
region. However, it ultimately leads to the unphysical
singularities of observables in the infrared domain. The Analytic
Perturbation Theory constitutes the next step of the improvement
of perturbative expansions. Specifically, it involves additional
analyticity requirement which is based on the causality principle
and implemented in the K\"allen--Lehmann and Jost--Lehmann
representations. Eventually, this approach eliminates spurious
singularities of the perturbative power series and enhances the
stability of the latter with respect to both higher loop
corrections and the choice of the renormalization scheme. The
paper contains an overview of the basic stages of the development
of the Analytic Perturbation Theory in QCD, including its recent
applications to the description of hadronic processes.
\end{abstract}

\maketitle

\section*{Preamble}

The method of the Analytic Perturbation Theory (APT) resolves the
problem of unphysical (or ghost) singularities of both the invariant
charge of Quantum Chromodynamics (QCD) and the matrix elements of the
strong interaction processes. This difficulty (known also as the
problem of Moscow zero or Landau pole) first appeared in Quantum
Electrodynamics (QED) in the mid-50s of the last century. It
played a certain dramatic role in the development of Quantum Field
Theory~(QFT).

In the late-50s Bogoliubov, Logunov, and Shirkov suggested~\cite{1}
resolving this problem by merging the renormalization group (RG)
method with the K\"allen--Lehmann representation, which implies the
analyticity in the complex $Q^2$-variable. The method of APT in QCD
is based on the ideas of Ref.~\cite{1}.

The development of the APT over the last decade has revealed a number
of new principal features of the analytic approach. Specifically, in
addition to the resolution of the problem of unphysical
singularities, the APT leads to the nonpower functional expansion for
QCD observables. The latter possesses an astonishing (in comparison
with the perturbative power series) stability with respect to both
higher loop corrections and the choice of the renormalization
prescription.

\section{Introduction}\label{sec1}

The renormalization group method, which was devised in the mid-50s in
Ref.~\cite{2} (see also paper~\cite{3} and the respective chapter of
book~\cite{4}), is an inherent part of the contemporary QFT
calculations. This method becomes especially useful for singular
solutions, when the type of the singularity is affected by
perturbative expansion. Besides, the RG method is important for
strong interactions, e.g., for~QCD.

The QCD description of the majority of hadronic processes
requires the use of the RG method. At the same time, the
straightforward solutions of the RG equations suffer from the
spurious singularities. The one-loop QCD invariant charge
$\bar\alpha_{\sa}(Q^2)$ possesses the ghost pole at $Q^2=\Lambda^2$,
see Eq.~\eqref{2.1} below. Higher loop corrections just give rise
to additional singularities of the cut type and do not eliminate
this problem. The existence of such singularities contradicts the
general principles of the local QFT~\cite{4},~\cite{5}.

A solution to the problem of the unphysical singularities of the
invariant charge was proposed in Ref.~\cite{1}. Specifically,
this can be achieved by merging the RG method with definite
properties of the analyticity in $Q^2$-variable. The latter follows
from the K\"allen--Lehmann spectral representation for the transverse
Lorentz-invariant amplitude of the dressed photon or boson
propagator
\begin{equation}
d(Q^2)=\frac{1}{\pi}\int_{0}^{\infty}
\frac{d\sigma}{\sigma+Q^2}\rho(\sigma),
\label{1.1}
\end{equation}
that reflects the basic principles of the local QFT.

In QED, the square of the electron effective charge
$\bar{\alpha}(Q^2)$, which was first introduced
by Dirac~\cite{6}, is proportional to the transverse
amplitude of the dressed photon propagator. The latter
satisfies the spectral K\"allen representation~\eqref{1.1},
which implies the analyticity in the complex $Q^2$-plane
with the cut along the negative semiaxis of real~$Q^2$.
The function $\bar{\alpha}(Q^2)$ is also called the invariant
charge or running coupling constant\footnote{We shall not
use the term ``running coupling constant'' here due to
its semantic nonsense.}. In accordance with paper~\cite{1},
the analytic invariant charge can be reconstructed by making
use of the representation~\eqref{1.1}, the spectral density
$\rho(\sigma)$ being defined as the discontinuity of the
perturbative QCD invariant charge across the physical cut
along the real negative semiaxis $\Req Q^2<0$.

The QED analytic invariant charge $\bar\alpha$, elaborated
in Ref.~\cite{1}, possesses the following important
properties:

-- it has no ghost pole;

-- as a function of $\alpha$ it has an essential singularity of the
form $\exp(-3\pi/\alpha)$ at the origin;

-- for real and positive $\alpha$ it admits a power expansion, that
coincides with the perturbative one;

-- $\bar\alpha$ has finite ultraviolet limiting value $3\pi$, which
is independent of the experimental value $\alpha\simeq1/137$.

In the mid-90s this idea was employed in QCD in
Refs.~\cite{7},~\cite{8}. Afterwards, the method developed therein
was named APT. In QCD, the synthesis of the renormalization
invariance with the $Q^2$-analyticity has revealed a number of
important features of the analytic invariant
charge~\cite{7},~\cite{8}. In particular, $\alpha_{\Ea}(Q^2)$ has the
{\it universal} infrared (IR) stable point. Its value
$\alpha_{\Ea}(Q^2=0)$ is determined by the one-loop $\beta$-function
coefficient $\beta_0$. The value $\alpha_{\Ea}(Q^2=0)$ is a
scheme-independent quantity, since it is not altered by multi-loop
corrections. The IR limiting value does not depend on the scale
parameter~$\Lambda$, which can be evaluated by making use of
experimental data. The set of curves $\alpha_{\Ea}(Q^2)$,
corresponding to different values of $\Lambda$, forms a bundle with
the common point $\alpha_{\Ea}(0)=1/\beta_0$. Therefore, the
imposition of the analyticity requirement essentially modifies the IR
behavior of the analytic invariant charge.

Another important feature of the analytic approach is that it enables
one to define the invariant coupling $\alpha_{\Ma}(s)$ in the
timelike (Minkowskian) domain in a self-consistent
way~\cite{9}. Usually, contemporary QFT calculations involve explicit
expressions for observables and other auxiliary RG-invariant (or
covariant) quantities, being expressed in terms of the invariant
charge. To achieve it, the quantity at hand has first to be
represented in an appropriate form. For example, only the quantity
defined in the Euclidean region can be expressed in terms of
$\bar\alpha_{\sa}(Q^2)$.

Meanwhile, a number of observables (e.g., the effective
cross-sections and quantities related to inclusive decays) are
functions of the timelike argument $s=-Q^2$, with $s$ being the
center-of-mass energy squared. However, in the framework of the RG
method, one cannot straightforwardly substitute the spacelike
(Euclidean) argument with the timelike (Minkowskian) one. Indeed, in
accordance with~\eqref{1.1} the amplitudes of propagators, similarly to
the transverse photon amplitude~$d(Q^2)$, and the relevant matrix
elements acquire complex values for real negative $Q^2$ (i.e., for
the timelike argument).

Interrelations between the RG-invariant quantities in the Euclidean
and Minkowskian domains can only be established by making use of the
linear integral transformations. The analytic properties of the
invariant charge, which are violated by perturbation theory (PT), and
can be recovered within the analytic approach afterwards, play a
crucial role here.

Both, the Minkowskian $\alpha_{\Ma}(s)$ and Euclidean
$\alpha_{\Ea}(Q^2)$ charges possess the common IR stable point
$\alpha_{\Ma}(s=0)= \alpha_{\Ea}(Q^2=0) = 1/\beta_0$. In the
ultraviolet (UV) region these couplings also have the same asymptotic
behavior. However, functions $\alpha_{\Ea}(Q^2)$ and
$\alpha_{\Ma}(s)$ cannot be identical in the entire energy range due
to certain general arguments~\cite{10}.

A key feature of APT is the transformation of series in powers of
${\bar{\alpha}_{\sa}}$ for observables into the {\it nonpower}
functional expansions. The latter displays both a milder dependence
on the choice of the renormalization scheme and an improved numerical
convergence.

Thus, one arrives at the self-consistent method of description of
observables. This approach possesses the renormalization invariance
and is free of unphysical singularities and related difficulties.

\section{Analytic perturbation theory}
\label{sec2}

\subsection{Singularities of the effective charge}
\label{ss2.1}

At the one-loop level, the RG summation of the ultraviolet
logarithms leads to the singular expression for the QCD running
coupling (``invariant charge'')
\begin{equation}
\bar\alpha_{\sa}^{(\ell=1)}(Q^2)=
\frac{\alpha_{\mu}}{1+\alpha_{\mu}\beta_0\ln(Q^2/\mu^2)}=
\frac{1}{\beta_0\ln(Q^2/\Lambda^2)},\qquad
\beta_0(n_f)=\frac{11-2n_f/3}{4\pi},
\label{2.1}
\end{equation}
where $n_f$ denotes the number of active flavors. The scale parameter
is defined here in the well-known way, namely $\Lambda=\mu
e^{-\frac{1}{2\alpha_{\mu}\beta_0}}$.

The function~\eqref{2.1} has the IR unphysical singularity at
$Q^2=\Lambda^2$. This problem cannot be solved~\cite{11} by taking
into account higher-loop corrections. Indeed, the latter just
modify the type of singularities, but do not eliminate them. For
example, the common form of the two-loop perturbative coupling
\begin{equation}
\bar\alpha_{\sa}^{(2)}(Q^2)=\frac{1}{\beta_0l}
\biggl[1-\frac{\beta_1}{\beta_0^2}\frac{\ln l}{l}\biggr]+
O\biggl(\frac{\ln^2l}{l^3}\biggr),\qquad
l=\ln\frac{Q^2}{\Lambda^2}
\label{2.2}
\end{equation}
possesses an additional unphysical cut due to the double-log
dependence on~$Q^2$.

A similar situation takes place in QED as well. It is worthwhile to
note here that in this latter case unphysical singularities
correspond to huge energy scales which have no physical meaning. At
the same time, in the case of QCD the value of the parameter
$\Lambda$ is about several hundred MeV, that is within the
physically-accessible range of energies.

\subsection{Analytic approach}\label{ss2.2}
In general, the renormalized perturbative expansion can be further
modified in a certain way. For example, the Adler function $D(Q^2)$
is representable by the double series in powers of the running
coupling $\alpha_{\mu}$ at a normalization scale $\mu^2$ and in
powers of $\ln(Q^2/\mu^2)$:
\begin{equation}\label{2.3}
D_{\pt}\biggl(\frac{Q^2}{\mu^2},\alpha_{\mu}\biggr)=\sum_{n}
\alpha_{\mu}^n\sum_{k=0}^nd_{n,k}\ln^k\biggl(\frac{Q^2}{\mu^2}\biggr).
\end{equation}
In the UV domain this expression is ill-defined due to a large value of
logs\footnote{This series is also meaningless in the IR domain due to
the same reason. Besides, one might expect the factorial growth of
the expansion coefficients $d_{n,k}$ at large~$n$.}.

The RG method improves the properties of the perturbative expansion
in the UV region by accumulating ``large logs'' into the invariant
charge
$\bar\alpha_{\sa}(Q^2)=\bar\alpha_{\sa}(Q^2/\mu^2,\alpha_{\mu})$. In
particular, the RG-improved Adler $D$-function becomes a function of
$\bar\alpha_{\sa}$ only, and can be represented by the power series
\begin{equation}
D_{\PT}(Q^2)=D\bigl(\bar\alpha_{\sa}(Q^2)\bigr)=
\sum_{n}d_{n,0}\bar\alpha_{\sa}^n(Q^2);\qquad
\text{{\rm PT}}=\text{{\rm pt}}+\text{{\rm RG}}.
\label{2.4}
\end{equation}

Unlike Eq.~\eqref{2.3}, the terms of Eq.~\eqref{2.4} are
$\mu$-independent, and the expansion parameter
$\bar\alpha_{\sa}(Q^2)$ vanishes when $Q^2/\mu^2\to\infty$, in
agreement with the asymptotic freedom.

At the same time, expansion~\eqref{2.4} formally remains ill-defined
in the IR domain due to unphysical singularities of the expansion
parameter~$\bar\alpha_{\sa}(Q^2)$. Thus, on the one hand, the RG
improvement of the perturbative series~\eqref{2.3} results in a crucial
physical property of the asymptotic freedom. On the other hand, an
important property of $D$-function, namely, the analyticity in the
complex $Q^2$-plane with the cut along the negative semiaxis of real
$Q^2$, is lost in Eq.~\eqref{2.4}.

The APT method constitutes the next step in the improvement of the
perturbative expansion. This method employs the principle of {\it
renormalization invariance} together with a fundamental principle of
{\it causality} which is realized in the form of the K\"allen--Lehmann
integral representation~\eqref{1.1}. In the framework of the APT, the
$D$-function can be represented by the nonpower functional expansion
\begin{equation}
D_{\PT}(Q^2)=\sum_{n}d_n\bar\alpha_{\sa}^n(Q^2)\quad\to\quad
D_{\APT}(Q^2)=\sum_{n}d_n{\mathcal{A}}_n(Q^2).
\label{2.5}
\end{equation}
The Euclidean functions ${\mathcal{A}}_n(Q^2)$ satisfy
the K\"allen--Lehmann representation
\begin{equation}
{\mathcal{A}}_n(Q^2)=\frac{1}{\pi}\int_0^{\infty}
d\sigma\,\frac{\rho_n(\sigma)}{\sigma+Q^2},
\label{2.6}
\end{equation}
with the spectral function being defined as the discontinuity of the
respective power of the invariant charge across the physical cut:
$\rho_k(\sigma)=\Ima\bar\alpha_{\sa}^k(-\sigma-i\epsilon)$. The
first-order function $\mathcal{A}_1(Q^2)$ corresponds to the
analytic running coupling
\begin{equation}
\alpha_{\Ea}(Q^2)=\frac{1}{\pi}\int_0^{\infty}
d\sigma\,\frac{\rho(\sigma)}{\sigma+Q^2},\qquad
\rho(\sigma)\equiv\rho_1(\sigma).
\label{2.7}
\end{equation}

The representation~\eqref{2.6} determines the analytic properties of
the functions ${\mathcal{A}}_n(Q^2)$ in the complex $Q^2$-plane.
Specifically, these functions (and, therefore, the
series~\eqref{2.5}), are analytic functions in the complex
$Q^2$-plane with the cut along the negative semiaxis of real~$Q^2$.
The investigation of the properties of these functions has revealed
that the resolution of the ghost pole problem by making use of the
APT method eventually leads to the IR stability with respect to
higher loop corrections. The results of this investigation are
summarized in Table~\ref{t1} which elucidates the stages of the
evolution of the perturbative results: PT $\to$ PT$+$RG $\to$ APT.

\begin{table}[htb]
\caption{Properties of various approximations}
\label{t1}
\smallskip
\begin{center}
\begin{tabular}{|c|c|c|c|c|}\hline
{{\small Method}}\vphantom{$\int^b$}& {\small{Type of approximation}}
&\multicolumn{3}{|c|}{\small{Properties}}
\cr
\cline{3-5}
&\vphantom{$\int^b$}&{\small{UV}} &{\small{IR}} &
{\small{Analyticity}}
\cr
\hline
{\small PT}\vphantom{$\int$} & {\small Double set in powers of} & & &
\cr
& $\alpha_{\mu}$ and $\ln Q^2/\mu^2$ & $-$ & $-$ & $+$
\cr
\hline
{\small PT} $+$ RG\vphantom{$\int$}& {\small Power series in} & & &
\cr
& {\small invariant charge} $\bar\alpha_{\sa}(Q^2)$& $+$& $-$ & $-$
\cr
\hline
{\small APT} $=$ {\small PT} $+$ {\small RG} $+$\vphantom{$\int_a^c$}
& {\small Nonpower functional expansions} & & &
\cr
$+$ {\small analyticity}
& {\small in} $\mathcal A_k(Q^2)$ {\small and} $\mathfrak A_k(s)$
&$+$ & $+$ & $+$
\cr
\hline
\end{tabular}
\end{center}
\end{table}

The last row of the Table contains two sets of functions, namely,
$\mathcal A_n(Q^2)$ and $\mathfrak{A}_n(s)$. The latter appear in the
description of the processes depending on the timelike (i.e.,
Minkowskian) momenta. These functions naturally emerge in the study
of the Drell function $R(s)$ which is the ratio of the inclusive
hadronic cross-section of the process of the $e^+e^-$-annihilation to
the leptonic one. Here the timelike argument $s$ is the center-of-mass
energy squared. The perturbative approximation of $R(s)$ by the power
series in the invariant charge,
$R_{\PT}(s)=\sum_{n}r_n\bar\alpha_{\sa}^n(s)$, violates the relation
between the functions $D(Q^2)$ and $R(s)$
\begin{equation}
D(Q^2)=Q^2\int_0^{\infty} ds\,\frac{R(s)}{(s+Q^2)^2}.
\label{2.8}
\end{equation}
In the framework of the APT, the function $R(s)$ takes the form
of the functional expansion
\begin{equation}
R_{\APT}=\sum_{n}d_n\mathfrak{A}_n(s),
\label{2.9}
\end{equation}
with the coefficients $d_n$ being identical to those of
Eq.~\eqref{2.5}. The relation between $\mathcal A_k(Q^2)$ and
$\mathfrak A_k(s)$
\begin{equation}
\mathcal A_k(Q^2)=Q^2\int_0^{\infty}ds\,
\frac{\mathfrak A_k(s)}{(s+Q^2)^2},
\label{2.10}
\end{equation}
is similar to the relation~\eqref{2.8} between $D(Q^2)$ and $R(s)$.
The functions $\mathfrak{A}_k$ can also be
expressed\footnote{Schwinger argued~\cite{12} that in QED the RG
$\beta$-function is proportional to the spectral function of the
photon propagator (i.e., invariant charge). However, this hypothesis
is violated beyond the two-loop level. In the framework of the APT,
the Schwinger's assumption turns out to be realized in a ``hybrid
form'' for the $\beta$-function corresponding to the Minkowskian
charge~\eqref{2.13}. Indeed, the logarithmic derivative of the latter
is proportional to the spectral function of the Euclidean charge, see
Eq.~\eqref{2.11}.} in terms of the spectral function
$\rho_k(\sigma)$~\cite{9}:
\begin{equation}
\mathfrak{A}_k(s)=\frac{1}{\pi}\int_{\sa}^{\infty}
\frac{d\sigma}{\sigma}\rho_k(\sigma).
\label{2.11}
\end{equation}

\begin{figure}[t]
\begin{center}
\includegraphics{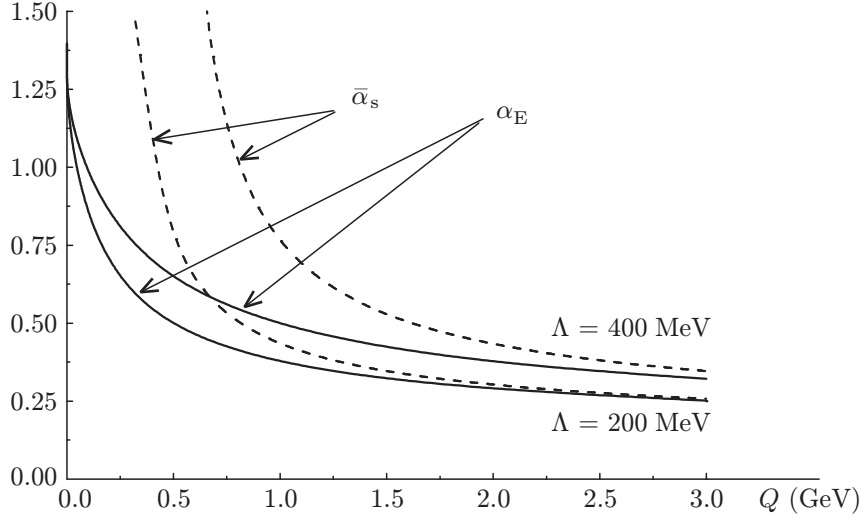}
\caption{Euclidean $\alpha_{\Ea}$ and perturbative $\bar\alpha_{\sa}$
charges.}
\label{fig1}
\end{center}
\end{figure}

\begin{figure}[t]
\begin{center}
\includegraphics{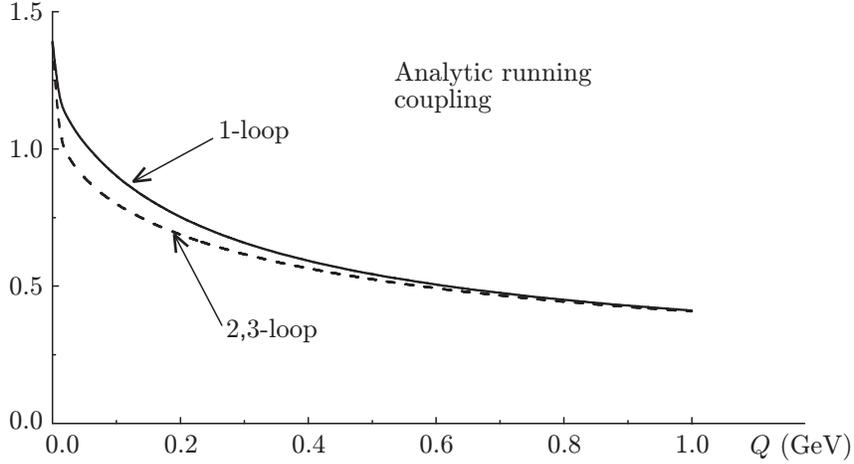}
\caption{Higher loop stability of the Euclidean charge
$\alpha_{\Ea}$.}
\label{fig2}
\end{center}
\end{figure}

In the leading order the Euclidean running coupling reads~\cite{7}
\begin{equation}
\alpha_{\Ea}^{(1)}(Q^2)=\frac{1}{\beta_0}
\biggl[\frac{1}{\ln(Q^2/\Lambda^2)}+
\frac{\Lambda^2}{\Lambda^2-Q^2}\biggr].
\label{2.12}
\end{equation}
Figure~\ref{fig1} depicts its behavior for $\Lambda=200$\,MeV and
$\Lambda=400$\,MeV. For comparison, the corresponding perturbative
curves are also plotted therein. The enhanced stability of the APT
expressions (in comparison with the perturbative ones) is demonstrated
in Fig.~\ref{fig2}, where the one-loop and two-loop functions are
shown. The three-loop analytic running coupling practically coincides
with the two-loop one (within the accuracy of 1--2\%).

Thus, contrary to the RG-improved perturbation theory, the
analyticity, which emerges from the causality, leads to the
stabilization of the behavior of the invariant charge in the IR
domain. A key property of the approach at hand is that all the
expansion functions assume the universal value at $Q^2=0$ which
eventually results in the stabilization mentioned above. At the same
time, the stability in the UV domain (starting from the two-loop
level) is due to the asymptotic freedom.

In the framework of the APT the invariant charge in the Minkowskian
region can be defined~\cite{13} in a self-consistent way as the first
of the functions~\eqref{2.11}:
\begin{equation}
\alpha_{\Ma}(s)=\frac{1}{\pi}\int_{\sa}^{\infty}
\frac{d\sigma}{\sigma}\rho(\sigma),\qquad
\rho(\sigma)\equiv\rho_1(\sigma).
\label{2.13}
\end{equation}
In the leading order this function reads
\begin{equation}
\alpha_{\Ma}^{(1)}(s)=\frac{1}{\beta_0\pi}
\arccos\frac{L}{\sqrt{L^2+\pi^2}}\biggr|_{L>0}=
\frac{1}{\beta_0\pi}\arctan\frac{\pi}{L},\qquad
L=\ln\frac{s}{\Lambda^2}.
\label{2.14}
\end{equation}

The Euclidean~\eqref{2.7} and Minkowskian~\eqref{2.13} charges share
the same IR limiting value
\begin{equation}
\alpha_{\Ea}(0)=\alpha_{\Ma}(0)=\frac{1}{\beta_0},
\label{2.15}
\end{equation}
which is independent of the loop level.

\subsection{Higher APT expansion functions}
\label{ss2.3}

These functions are necessary for the analysis of observables by
making use of Eqs.~\eqref{2.9} and~\eqref{2.5}. They satisfy
the recurrent relations
\begin{equation}
\frac{1}{k}\frac{d\mathfrak A_k(s)}{d\ln s}=
-\sum_{n\geq1}\beta_{n-1}\mathfrak A_{k+n}(s),\qquad
\frac{1}{k}\frac{d\mathcal A_k(Q^2)}{d\ln Q^2}=
-\sum_{n\geq1}\beta_{n-1}\mathcal A_{k+n}(Q^2),
\label{2.16}
\end{equation}
which can be employed for their iterative definitions. To achieve it,
one has to explicitly solve these relations by making use of
additional assumptions of the form $\beta_{\ell\geq\ell_m}=0$ and
$A_{k\geq K+1}=0$. At the one-loop level ($\beta_{\ell\geq1}=0$) the
APT formulae have a simple and elegant form. In this case, proceeding
from the first functions~\eqref{2.12} and~\eqref{2.14}, one can show
by making use of relations~\eqref{2.16} that
\begin{equation}
\begin{alignedat}3
&\mathcal A_2^{(1)}(l)=\frac{1}{\beta_0^2}
\biggl(\frac{1}{l^2}-\frac{e^l}{(e^l-1)^2}\biggr),&\quad
&\mathfrak A_2^{(1)}(L)=\frac{1}{\beta_{0}^2}\frac{1}{L^2+\pi^2},&\quad
&l=\ln\biggl(\frac{Q^2}{\Lambda^2}\biggr),
\\
&\mathcal A_3^{(1)}(l)=\frac{1}{\beta_0^3}\biggl(\frac{1}{l^3}-
\frac{1}{2}\frac{e^l+e^{2l}}{(e^l-1)^3}\biggr),&\quad
&\mathfrak A_3^{(1)}(L)=\frac{1}{\beta_{0}^3}\frac{L}{(L^2+\pi^2)^2},
&\quad
&L=\ln\biggl(\frac{s}{\Lambda^2}\biggr).
\end{alignedat}
\label{2.17}
\end{equation}

The two-loop level is more complicated technically. The point is that
the exact solution for $\alpha_{\sa}$ is expressed in terms of the
special Lambert function here, which leads to cumbersome explicit
expressions for $\mathcal A_k$ and $\mathfrak A_k$~\cite{14}.

Nonetheless, all APT functions obey the following important
properties at any loop level:

$\bullet$ Unphysical singularities are absent, no additional
parameters being introduced.

$\bullet$ Higher functions,~\eqref{2.17} etc., are not equal to
powers of the first ones~\eqref{2.12},~\eqref{2.14}. They oscillate
in the vicinity of $|Q^2|\sim\Lambda^2$ and vanish in the IR limit,
see Fig.~\ref{fig3}b below. At the same time, these functions tend to
the powers of~$\alpha_{\sa}$ in the UV asymptotic.

$\bullet$ The expansions of observables in powers of
$\alpha_{\sa}(Q^2)$ (for the Euclidean case) and in powers of
$\alpha_{\sa}(s)$ (for the Minkowskian case) are replaced by the
expansions over the sets of $\bigl\{\mathcal A_k(Q^2)\bigr\}$ and
$\bigl\{\mathfrak A_k(s)\bigr\}$, respectively. The latter expansions
display a faster convergence with respect to that of the perturbative
case.

\subsection{Analyticity in the $\alpha_{\sa}$-plane}
\label{ss2.4}

As it has been noted above, the expression for the Euclidean coupling
$\alpha^{(1)}_{\Ea}(Q^2/\mu^2,\alpha_{\mu})$ contains the nonanalytic
term of the form $\exp[-1/\alpha_{\sa}]$. The latter corresponds to
the essential singularity at the origin of the
complex~$\alpha_{\sa}$-plane.

More than half a century ago, proceeding from a general reasoning,
Dyson argued~\cite{15} that such singularity appears in QED
inevitably. The explicit form of this singularity coincides with both
the term determined in Ref.~\cite{16} by merging the renormalization
invariance with the causality condition and with the results of
Ref.~\cite{17} obtained by making use of the functional saddle-point
method.

It is worth noting also that the logical inevitability of the
nonpower type of the functional APT-expansions was discussed in
detail in Refs.~\cite{18}.

The conversion of the common QCD running
coupling $\alpha_{\sa}(L)$ into the
Euclidean $\alpha_{\Ea}(L)$ or Minkowskian $\alpha_{\Ma}(L)$ one is
equivalent to the introduction of a new expansion parameter. It is
worth considering two examples
\begin{equation}
\alpha\to w_1^{\Ma}(\alpha)=\frac{1}{\pi\beta_0}
\arccos\frac{1}{\sqrt{1+\pi^2\beta_0^2\alpha^2}},\qquad
\alpha\to w_1^{\Ea}(\alpha)=\alpha+\frac{1}{\beta_0}
(1-e^{1/(\beta_0\alpha)})^{-1}.
\label{2.18}
\end{equation}
The first one is similar to the choice of another renormalization
scheme, since the function $w^{\mathrm{M}}(\alpha)$ can be expanded
in powers of~$\alpha$. The other one transforms into identity in the
weak coupling limit, since its second term $e^{-1/(\beta_0\alpha)}$
does not contribute to the expansion in powers of $\alpha$ at $\alpha
\to 0$.

At the same time, the conversions~\eqref{2.18} give rise to the
transformations of the invariant charge
$\alpha_{\sa}^{(1)}(L)\to{\mathfrak A}_1^{(1)}(L)$ and
$\alpha_{\sa}^{(1)}(L)\to\mathcal A_1^{(1)}(L)$ which are equivalent
to Eqs.~\eqref{2.12} and~\eqref{2.14}.

The function $w^{\mathrm{E}}_1(\alpha_{\sa})$ possesses an essential
singularity at the origin of the complex $\alpha_{\sa}$-plane which
agrees with the results of Refs.~\cite{15}--\cite{17}.

Similarly, the sets of higher-order functions of the APT
expansions, $\bigl\{\mathfrak A_k(x)\bigr\}$, $\bigl\{\mathcal
A_k(x)\bigr\},\dots$ map onto the nonpower sets
$\bigl\{w^{\Ma}_k(\alpha_{\sa})\bigr\},
\bigl\{w^{\Ea}_k(\alpha_{\sa})\bigr\}, \dots$ of the
functions\footnote{Here $\aleph_k(\ln r^2\Lambda^2)$ denote the APT
functions in the configuration representation which are related to
$\mathcal A_k$ by the Fourier transformation~\cite{19}.}:
\begin{equation}
\begin{aligned}
&w^{\Ma}_k(\alpha_{\sa})=\mathfrak A_k
\biggl(L=\frac{1}{\alpha_{\sa}}\biggr),\qquad
w^{\Ea}_k(\alpha_{\sa})=\mathcal A_k\biggl(\frac{1}{\alpha_{\sa}}\biggr),
\\
&w^{\Da}_k(\alpha_{\sa})=\aleph_k\biggl(\ln r^2\Lambda^2=
\frac{1}{\alpha_{\sa}}\biggr),\quad\dots \quad.
\end{aligned}
\label{2.19}
\end{equation}
The adjacent elements satisfy simple differential relations
\begin{equation}
w^R_{k+1}(\alpha_{\sa})=\frac{\alpha_{\sa}^2}{k}
\frac{dw^R_k(\alpha_{\sa})}{d\alpha_{\sa}},\qquad
R=\Ea,\Ma,\Da,\dots,
\label{2.20}
\end{equation}
which follow from Eqs.~\eqref{2.16} at the one-loop level. Meanwhile,
the same-order functions from different sets are interrelated with
each other by the integral transformations following from
Eq.~\eqref{2.10}. Besides, all the Euclidean functions
$w^{\Ea}_k(\alpha)$ possess an essential singularity at $\alpha=0$.

The regularity of the behavior of the timelike APT-invariant charge
$\alpha_{\Ma}$ and its effective ``powers'' $\mathfrak A_k$ is
provided by the summation~\cite{20} of an infinite number of
perturbative contributions related to the $\pi^2$-terms. In turn,
the result of the $\pi^2$-summation in the Minkowskian region, being
extended into the Euclidean domain~\eqref{2.10}, leads to the
recovery of the nonanalytic contributions of the form
$\exp\bigl(-1/(\beta_0\alpha_{\mu})\bigr)$ which are ``invisible''
within the original perturbative expansion. Therefore, the Euclidean
functions $\mathcal A_k$ contain both logarithmic and power terms
in~$Q^2$.

Apparently, the sets of the functions
$\bigl\{w^R_k(\alpha_{\sa})\bigr\}$, appearing in the
``$\alpha_{\sa}$-representation'', are similar to Caprini--Fischer
sequences~\cite{21} which have been obtained proceeding from a
different reasoning.

\subsection{Global APT and the ``distorted'' mirror}
\label{ss2.5}

In the studies of the hadronic processes at various energy scales one
has to take into account the dependence of the theoretical results on
the active quark flavor number~$n_f$. Following the Bogoliubov method
of massive renormgroup, the algorithm of the smooth matching was
devised in Ref.~\cite{22}. This algorithm was employed for the
analysis of the evolution of the running coupling in the range
3\,GeV${}<Q<100$\,GeV~\cite{23}.

For the massless schemes, the matching of the invariant charge at the
``Euclidean thresholds'' is commonly employed~\cite{24}. However,
this procedure destroys the smoothness of the Euclidean function,
and, therefore, violates its analyticity.

\begin{figure}[h!]
\centerline{\includegraphics[scale=0.9]{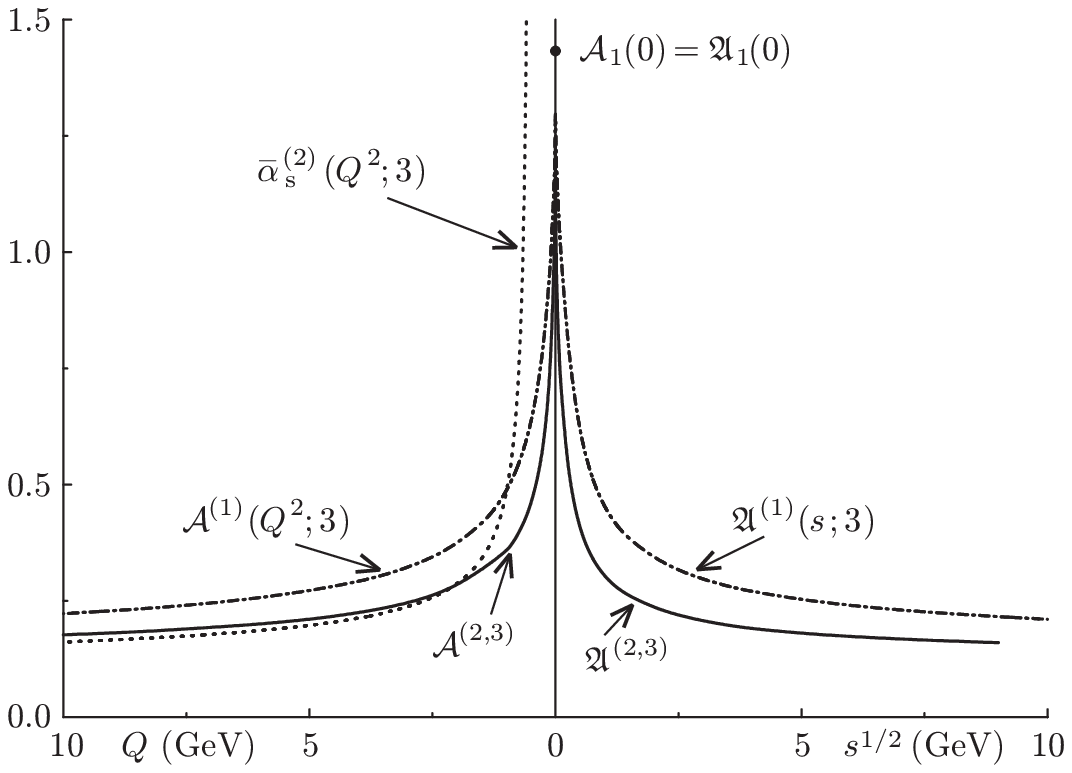}}
\centerline{\small a}
\bigskip
\centerline{\includegraphics[scale=0.9]{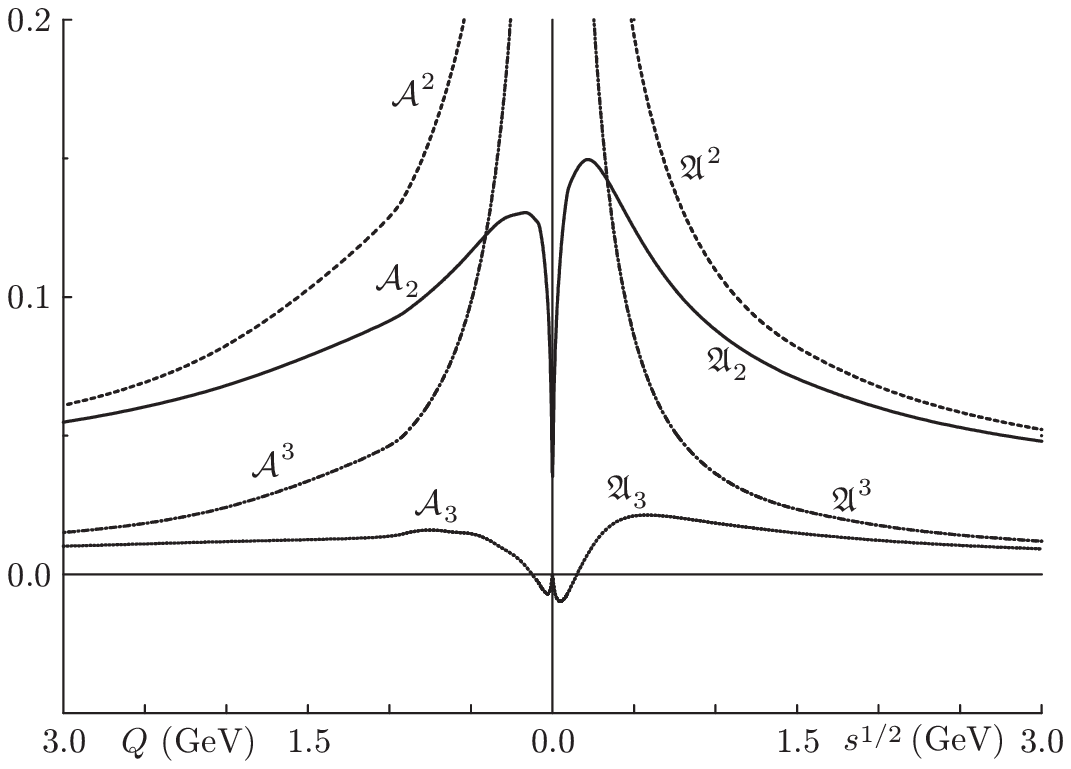}}
\centerline{\small b}
\caption{a) The global Euclidean and Minkowskian QCD
running couplings for $\Lambda^{(3)}=350$\,MeV; b)
the ``distorted mirror'' which elucidates the
asymmetry of Euclidean and Minkowskian functions.}
\label{fig3}
\end{figure}

The APT method opens a new opportunity for a self-consistent
description of the observables in domains corresponding to various
numbers of active quarks $n_f$ \cite{25},~\cite{26}. Proceeding from
the common matching condition, one defines the spectral functions
$\rho(\sigma,n_f)$ and employs them in Eqs.~\eqref{2.6}
and~\eqref{2.11} afterwards. This procedure provides the analyticity
of the Euclidean functions, whereas the Minkowskian functions turn
out to be the piecewise smooth ones. Eventually, this results in the
``global'' APT-functions which ``know'' about all quark thresholds.

Figure~\ref{fig3}a, taken from Ref.~\cite{26}, depicts the global
APT-charge in the Euclidean and Minkowskian domains.
Figure~\ref{fig3}b demonstrates the ``distorted mirror'', i.e., that
the Euclidean and Minkowskian functions possess a similar, but
asymmetrical behavior. For comparison with higher functions
of the APT-expansions ${\mathcal{A}}_k(Q^2)$ and $\mathfrak A_k(s)$,
the curves corresponding to the powers of the first APT functions are
also given therein.

\subsection{Possible developments of the minimal APT}
\label{ss2.6}

As it has already been noted, the present version of the APT contains
no additional parameters in comparison with the common RG-improved
PT. We call it the {\it minimal} one. The straightforward application
of the minimal APT in the low energy region (i.e., for the energies
of the order of $\Lambda$) is not indisputable. It is worth noting
that at such energies the effects due to the quark masses become
considerable. In this situation, it is natural to modify the minimal
APT by introducing new parameters. It is worthwhile to mention the
so-called ``synthetic'' modification explored in Ref.~\cite{27}. Here
the running coupling acquires the IR enhancement controlled by an
additional parameter. In turn, this allows one to establish a link
with the potential quark model. Other variants of the APT involve
effective parton masses or modify the spectral representation by
shifting the lower integration limit to the two-pion
threshold~\cite{28}. Besides, there are more formal ways of
modification of the minimal APT at low energies~\cite{29}.

It is worthwhile to note that for numerical estimations one may
employ the results of Ref.~\cite{14} which also contains the
expressions for the invariant expansion functions in terms of the
Lambert function. Since these expansions are rather cumbersome, for
practical applications it proves to be convenient to employ simple
approximate formulae. A simple explicit expression for the Euclidean
QCD function was first proposed in Ref.~\cite{30} in the form of
a one-parameter model
\begin{equation}
{\mathcal{A}}_{\bmod}^{(2)}(Q^2)=\frac{1}{\beta_0}
\biggl[\frac{1}{l_2(a)}+\frac{1}{1-\exp\bigl(l_2(a)\bigr)}\biggr].
\label{2.21}
\end{equation}
This model is based on the one-loop expression with the
modified argument
\begin{equation}
l_2=l+\frac{\beta_1}{2\beta_0^2}\ln(l^2+a\pi^2),\qquad
l=\ln\frac{Q^2}{\Lambda^2},
\label{2.22}
\end{equation}
where $a=4$. Recently this model has been essentially developed. It
was shown~\cite{31}, that the other choice of parameter $a=2$ allows
one to approximate both the Euclidean and Minkowskian three-loop APT
functions $\mathcal A_k$, $\mathfrak A_k$, $k=1,2,3$ within the
accuracy sufficient for the description of all the experimental data
above 1\,GeV. The recurrent relations, modified in a proper way, lead
to simple expressions of the ``one-loop'' form~\eqref{2.17} for
higher functions. This enables one to employ the new model based on
the substitution~\eqref{2.22} in the analysis of contemporary
experimental data without technical difficulties. This model has been
used in the analysis of the inclusive $\Upsilon$-decay in
Ref.~\cite{31}. It was revealed therein that a weak point of the
theoretical processing of rather precise experimental data is the
choice of the scale.

\section{Phenomenological applications}
\label{sec3}

The analytic approach has been successfully employed in studies of
many hadronic processes. The literature devoted to the applications
of the APT method is rather vast. Below, we mention some important
results.

\subsection{Inclusive $\tau$-lepton decay}
\label{ss3.1}

The $\tau$~lepton is the only lepton which is heavy enough to decay
into hadrons. Experimental data on the inclusive $\tau$~lepton
decay into hadrons possess good accuracy in comparison with those
of other hadronic processes. These data constitute a ``natural
ground'' for testing the low-energy QCD.

The importance of the analyticity in the description of the
$\tau$~decay can be elucidated by the following example. The
experimentally measurable quantity $R_{\tau}$ is related to the
life-time of the $\tau$~lepton. The accuracy of its measurement is
about 1\,\%. At the same time, $R_{\tau}$ can be represented as the
integral of the imaginary part of the correlation function
\begin{equation}
R_{\tau}=\frac{2}{\pi}\int^{M_{\tau}^2}_0\frac{ds}{M_{\tau}^2}
\biggl(1-\frac{s}{M_{\tau}^2}\biggr)^2
\biggl(1+2\frac{s}{M_{\tau}^2}\biggr)\Ima\Pi(s).
\label{3.1}
\end{equation}
The principal difficulty of the theoretical analysis of $R_{\tau}$ is
due to the fact that the integration range in Eq.~\eqref{3.1}
involves the low energy region. Meanwhile, the standard perturbation
theory is not valid in the IR domain. Besides, if $\Ima\Pi(s)$ is
parameterized by the power series in~$\bar\alpha_{\sa}$, the integral
in Eq.~\eqref{3.1} does not exist. This latter difficulty can be
avoided by proceeding to the contour integral in the complex
$s$-plane
\begin{equation}
R_{\tau}=\frac{1}{2\pi{i}}\oint_{|s|=M_{\tau}^2}
\frac{ds}{s}\biggl(1-\frac{s}{M_{\tau}^2}\biggr)^3
\biggl(1+\frac{s}{M_{\tau}^2}\biggr)D(-s),
\label{3.2}
\end{equation}
where $D(-s)$ is the Adler function defined in Eq.~\eqref{2.8}.
However, the original Eq.~\eqref{3.1} can be represented in the
form of Eq.~\eqref{3.2} only if the correlator satisfies the required
analytic properties. This latter condition holds in the framework of
the APT. The inclusive $\tau$~lepton decay was studied in
Refs.~\cite{32},~\cite{33}.

\subsection{$e^+e^-$-annihilation into hadrons}
\label{ss3.2}

The process of $e^+e^-$-annihilation into hadrons was examined
in the framework of the APT in Ref.~\cite{34}. The measurable
quantity here is $R(s)$ which is the ratio of the hadronic
cross-section to the leptonic one. The theoretical analysis of $R(s)$
entails certain complications. The APT results presented below
correspond to the so-called ``smeared'' function
$R_{\Delta}(s)$~\cite{35}. The parameter~$\Delta$ specifies a minimal
``safe'' distance from the cut in the complex $s$-plane which
guarantees the absence of difficulties of the theoretical description
of resonances.

\begin{figure}[t]
\centering
\includegraphics{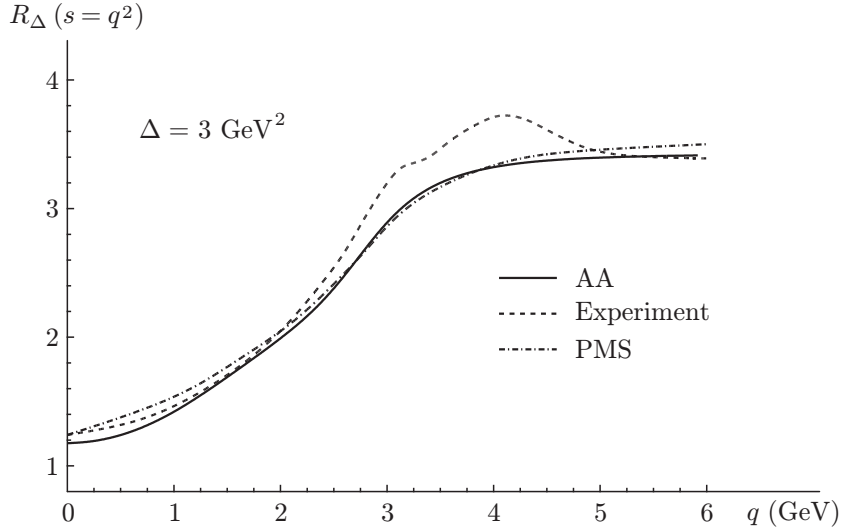}
\caption{The smeared ``experimental'' function $R_{\Delta}(q^2)$
vs.~the results of the APT method and the PMS-optimization.}
\label{fig4}
\end{figure}

The function $R_{\Delta}(s)$ reads
\begin{equation}
R_{\Delta}(s)=\frac{1}{2i}\bigl[\Pi(s+i\Delta)-\Pi(s-i\Delta)\bigr],
\label{3.3}
\end{equation}
with~$\Delta$ being a finite parameter. By making use of the
dispersion relation for~$\Pi(q^2)$, one can express the
function~\eqref{3.3} in terms of the measurable ratio~$R(s)$:
\begin{equation}
R_{\Delta}(s)=\frac{\Delta}{\pi}\int_0^{\infty} ds'\,
\frac{R(s')}{(s-s')^2+\Delta^2}.
\label{3.4}
\end{equation}
The function $R(s)$ in the integrand can be approximated by the
relevant experimental data at low and intermediate energies and by
its perturbative prediction at high energies. This allows one to
obtain the ``experimental'' curve for~$R_{\Delta}(s)$. The
integration of Eq.~\eqref{3.4} results in the smearing of the
resonance structure of~$R(s)$. The other quantity, which we use in
comparing the experimental and theoretical results, is the
mentioned-above Euclidean Adler $D$-function. The ``experimental''
curve for $D(Q^2)$ can be obtained in the same way as
for~$R_\Delta(s)$.

\begin{figure}[t]
\centering
\includegraphics{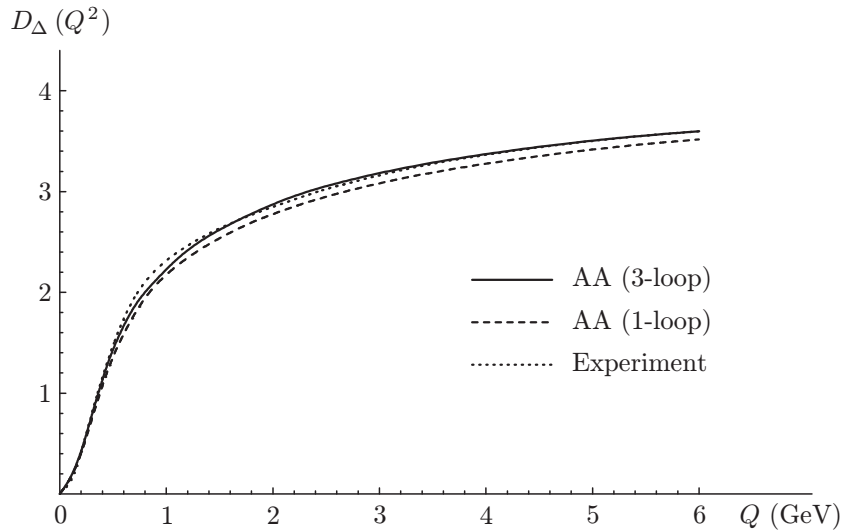}
\caption{The ``experimental'' function $D_{\Delta}(Q^2)$
vs.\ the APT results.}
\label{fig5}
\end{figure}

In Figs.~\ref{fig4} and~\ref{fig5}, the comparison of the APT results
(denoted by the ``AA'' labels) with the experimental prediction is
presented. Figure~\ref{fig4} also shows the curve found with the PMS
optimization~\cite{36} of the third-order perturbative expansion. The
experimental curve is also taken from Ref.~\cite{36}. It is worth
emphasizing that the APT method and PMS optimization originate in
different reasonings, though lead to close numerical results. Further
we show that, contrary to perturbation theory, the APT results
are practically independent of the subtraction scheme. Therefore,
additional optimization of the scheme dependence in APT is not as
important as in the perturbative case. The ``experimental''
curve for~$D_\Delta$ presented in Fig.~\ref{fig5} is taken from
Ref.~\cite{37}. Figures~\ref{fig4} and~\ref{fig5} demonstrate good
agreement of the theoretical APT results with the ``smeared''
phenomenological functions $R_\Delta$ and $D_\Delta$ which have been
reconstructed by making use of experimental data on the
$e^+e^-$-annihilation into hadrons.

\subsection{Renormalization scheme dependence}
\label{ss3.3}

An inevitable truncation of perturbative series leads to the
well-known problem of the renormalization scheme dependence. It is
worth noting that there is no firm criterion of the choice of the
renormalization prescription. The partial sum of the perturbative
series bears a dependence on the renormalization scheme, which is a
source of the theoretical ambiguity in processing the data. In QCD,
such ambiguity is the greater the smaller the energy scale is.
Therefore, the analysis of the stability of the results should
involve the investigation of both higher loop and scheme stability.

\begin{figure}[t]
\centering
\includegraphics{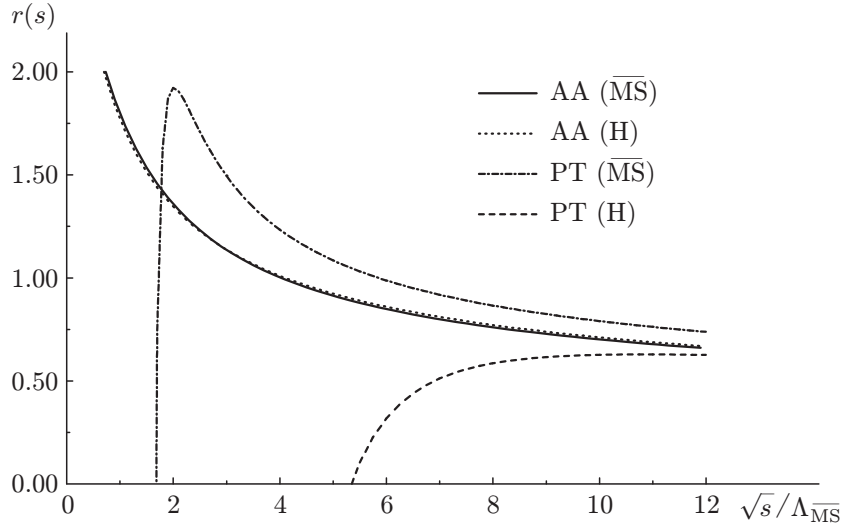}
\caption{The strong correction to the $R$-ratio within the analytic
approach (AA) and the standard perturbation theory (PT) for various
renormalization schemes.}
\label{fig6}
\end{figure}

The scheme stability of the APT results was first examined in
Ref.~\cite{38}. In Fig.~\ref{fig6} (taken from Ref.~\cite{38}), the
strong correction~$r(s)$ to the $R$-ratio of $e^+e^-$-annihilation
into hadrons $R(s)\propto1+r(s)$ is shown. The curves presented
therein were calculated within the APT and PT approaches at the
three-loop level. In these calculations the widely-accepted
$\overline{\mathrm{MS}}$-scheme and the so-called $\Ha$-scheme were
employed. The latter scheme is close to the former one in the sense
of the cancelation index. The issue of the scheme dependence was
discussed in detail in Ref.~\cite{30}.

Figure~\ref{fig6} shows that for the energy of about few GeV the
standard perturbative approach leads to large ambiguity due to the
scheme dependence. The APT method drastically reduces the scheme
dependence of the theoretical results. In particular, the APT curves
corresponding to different schemes practically coincide. A similar
result also takes place for the inclusive $\tau$~lepton
decay~\cite{33} and for the sum rules of the deep inelastic
lepton-hadron scattering~\cite{39}.

\subsection{Hadronic contribution to the muon anomalous
magnetic moment and to the fine structure constant}
\label{ss3.4}

The APT method has recently been applied~\cite{40} to the description
of the so-called $R$-related quantities. Among them the hadronic
contribution to the muon anomalous magnetic moment plays an important
role. The latter quantity (in the leading order in the
electromagnetic coupling $\alpha=\alpha_{\QED}$) reads
\begin{equation}
a_{\mu}^{\had}=
\frac{1}{3}\biggl(\frac{\alpha}{\pi}\biggr)^2
\int_{0}^{\infty}\frac{ds}{s}K(s)R(s),
\label{3.5}
\end{equation}
with $K(s)$ being a known function.

The hadronic contribution to the fine structure constant $\alpha$ can
be represented in the form
\begin{equation}
\Delta\alpha_{\had}^{(5)}(s)=-\frac{\alpha}{3\pi} s\:
{\mathcal{P}}\!\!\int_0^{\infty}\frac{ds'}{s'}
\frac{R(s')}{s'-s}.
\label{3.6}
\end{equation}
The superscript ``(5)'' implies that the contributions of only first
five quarks ($u$, $d$, $s$, $c$, and~$b$) was retained here.

Similar to the case of the $\tau$~decay, the integration range in
these expressions includes the low-energy region where the
perturbation theory is inapplicable. Quantities~\eqref{3.5}
and~\eqref{3.6} were evaluated in the framework of the APT in
Ref.~\cite{40}. The latter also employs the assumption about the
behavior of the quark mass function at low energy which is based on
the nonperturbative solution of the Schwinger--Dyson equation. The
obtained theoretical value
\begin{equation}
a_{\mu}^{\had}=(698\pm13)\times10^{-10}
\label{3.7}
\end{equation}
is in good agreement with the phenomenological estimations of
$a_{\mu}^{\had}$ which employ the data on $e^+e^-$-annihilation and
$\tau$~decay~\cite{41},~\cite{42}.

The hadronic contribution to the fine structure constant at the
$Z$-boson scale evaluated in Ref.~\cite{40}
\begin{equation}
\Delta\alpha_{\had}^{(5)}(M_Z^2)=(278.2\pm3.5)\times10^{-4}
\label{3.8}
\end{equation}
agrees with the estimation of Ref.~\cite{42} which uses the data on
$e^+e^-$-annihilation into hadrons
\begin{equation}
\Delta\alpha_{\had}^{(5)}(M_Z^2)=
(275.5\pm1.9_{\expt}\pm1.3_{\rad})\times10^{-4}.
\label{3.9}
\end{equation}

Besides, the method of APT provides a reasonable description of some
other quantities, e.g., the inclusive $\tau$~decay characteristic in
the vector channel~$R_{\tau}^{\Va}$, and the functions
$D_{\Delta}^{\tau}(Q^2)$ and $R_{\Delta}^{\tau}(s)$ corresponding to
the $\tau$~decay data.

\subsection{Some other applications}
\label{ss3.5}

\begin{figure}[t]
\centering
\includegraphics{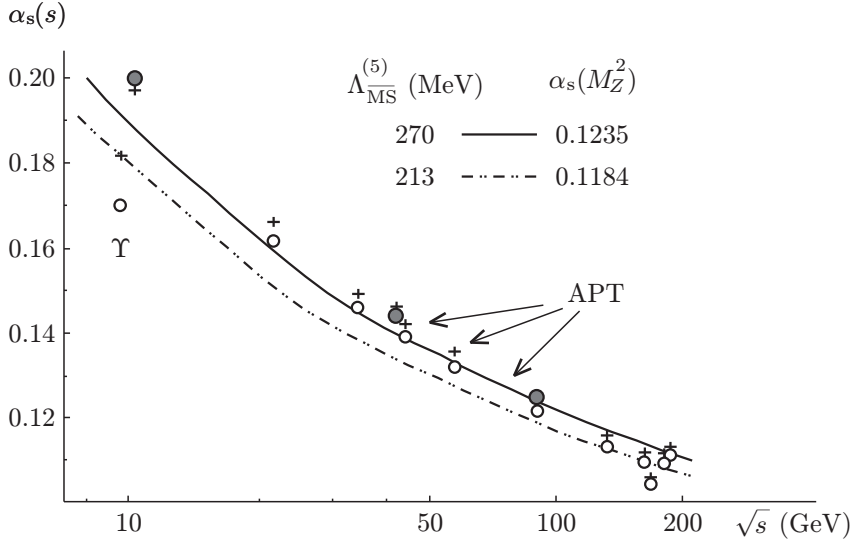}
\caption{The analysis of the five-quark timelike domain. The
difference between the points denoted by $+$ and $\circ$, $\bullet$
is due to the $\pi^2$-corrections. The solid APT curve corresponds to
$\Lambda_{\overline{\text{{\rm MS}}}}^{(5)}=270$\,MeV and
$\alpha_{\sa}(M_Z^2)=0.124$. The standard dot-dashed curve
corresponds to $\Lambda_{\overline{\text{{\rm MS}}}}^{(5)}=213$\,MeV
and $\alpha_{\sa}(M_Z^2)=0.118$.}
\label{fig7}
\end{figure}

In the framework of the APT the observables in the timelike domain
($s$--channel) can be represented as the nonpower expansion over the
functions which retain the so-called $\pi^2$-terms. The analysis of
the $s$--channel observables~\cite{26} has revealed the following.
For the energies above $50$\,GeV ($n_f=5$) the running coupling
$\alpha_{\sa}$ gains the effective positive shift
$\Delta{\bar{\alpha}_{\sa}} \simeq +0.002$ with respect to the
standard two-loop (NLO) analysis. In the energy range $10\div50$\,GeV
($n_f=5$) the value of this shift increases, namely,
$\Delta{\bar{\alpha}_{\sa}} \simeq +0.003$. This leads to a new
value of the QCD invariant charge at the scale of the $Z$-boson mass:
$\bar\alpha_{\sa}(M_Z^2)=0.124$. The obtained results are presented
in Fig.~\ref{fig7} taken form Ref.~\cite{26}.

The convergence of the APT expansions is better than that of the
perturbative series which is demonstrated in Table~\ref{t2}. The
results for the Gross--Llewellyn Smith sum rule for the deep
inelastic lepton-hadron scattering at $\sqrt{Q^2}\,\sim1.76$\,GeV
(Euclidean domain) and for the inclusive $\tau$~lepton decay
($M_{\tau}=1.777$\,GeV, timelike region) are given therein. Other
applications are presented in Table~2 of Ref.~\cite{26}.

\begin{table}[h!]
\caption{Various order contributions to observables
within PT and APT methods}
\label{t2}
\begin{center}
\begin{tabular}{|c|c|c|c|c|}\hline
Process&Method&1st order&2nd order&3rd order
\cr
\hline
Gross--Llewellyn Smith &PT & 65.1\% & 24.4\% &10.5\%
\cr
sum rule ($\sqrt{Q^2}\,\sim1.76$\,GeV) &APT & 75.7\% & 20.7\% &3.6\%
\cr
\hline
Inclusive $\tau$~lepton &PT & 54.7\% & 29.5\% &15.8\%
\cr
decay ($M_{\tau}=1.777$\,GeV) &APT & 87.9\% & 11.0\% &1.1\%
\cr
\hline
\end{tabular}
\end{center}
\end{table}

It is worthwhile to mention several other applications of APT. The
analytic running coupling has been successfully employed in the
analysis of the meson spectroscopy~\cite{43}. A recent (preliminary)
result obtained by the Milano group shows that the form of the QCD
interaction extracted from the light quarkonium spectrum as a
function of $Q^2$ below 1\,GeV can be well approximated by the
three-loop Euclidean~$\alpha_{\Ea}^{(3)}$, the value of the scale
parameter $\Lambda$ being close to its world average
value\footnote{We thank Prof.~G.Prosperi for providing us with this
information.}.

The Euclidean running coupling and the APT nonpower expansion were
employed in the description of the formfactor of the pion-photon
transition with Sudakov suppression~\cite{44} and the electromagnetic
formfactors~\cite{45}. Besides, a strong reduction of the sensitivity
of the results with respect to both the choice of the factorization
scale~\cite{46} and the renormalization scheme~\cite{47} was
revealed.

\section{Further development of the APT:
the inelastic lepton-hadron scattering}
\label{sec4}

In the previous sections, the APT method has been applied in the
analysis of the physical processes which can be described in terms of
the two-point function, namely, the correlator of the quark currents.
The analyticity condition was employed in the form of the
K\"allen--Lehmann representation. In this section, we address the
inelastic lepton-hadron scattering which can be characterized by the
structure functions of two scalar arguments. These functions possess
rather complicated analytic properties. Nonetheless, the basic idea
of the analytic approach turns out to be useful in this case, too.

\subsection{Jost--Lehmann representation}
\label{ss4.1}

In the case of the inelastic lepton-hadron scattering, the general
principles of the axiomatic QFT are accumulated in the Jost--Lehmann
(JL) representation\footnote{This representation is also known as
the Jost--Lehmann--Dyson representation, see~\cite{4}.}~\cite{48} for
the structure functions. In the nucleon rest frame this representation
reads~\cite{49}
\begin{equation}
W(\nu,Q^2)=\varepsilon(q_0)\int d{\mathbf u}\,d\lambda^2\,
\delta\bigl[q_0^2-(M{\mathbf u}-{\mathbf q})^2-\lambda^2\bigr]
\psi({\mathbf u},\lambda^2),
\label{4.1}
\end{equation}
the support of the distribution $\psi({\mathbf u},\lambda^2)$
being localized on the manifold
\begin{equation}
\rho=|{\mathbf u}|\leq1,\qquad\lambda^2\geq\lambda_{\min}^2=
M^2\bigl(1-\sqrt{1-\rho^2}\,\bigr)^2.
\label{4.2}
\end{equation}

For the physical process of inelastic scattering the variables $\nu$
and $Q^2$ assume positive values. It is convenient to introduce a
symmetric in $\nu$ function, to be denoted by the
same~$W(\nu,Q^2)$. By making use of the radial symmetry of the
distribution $\psi({\mathbf u},\lambda^2) = \psi(\rho,\lambda^2)$,
which follows from the covariance, one can rewrite the JL
representation in the covariant form~\cite{30}
\begin{equation}
W(\nu,Q^2)=\int_0^1d\rho\,\rho^2\int_{\lambda_{\min}^2}^{\infty}d\lambda^2
\int_{-1}^{1}dz\,\delta\bigl(Q^2+M^2\rho^2+\lambda^2-
2z\rho\sqrt{\nu^2+M^2Q^2}\,\bigr)\psi(\rho,\lambda^2).
\label{4.3}
\end{equation}

In what follows we adopt the common notation $Q^2=-q^2$ and
$\nu=q{P}$, where $P$ is the hadron momentum and $q$ stands for the
momentum transferred.

\subsection{Dispersion relation for the forward scattering amplitude}
\label{ss4.2}

Proceeding from the representation~\eqref{4.3}, one can derive the
$\nu$-dispersion relation (DR) for the virtual forward Compton
scattering amplitude~$T(\nu,Q^2)$
\begin{equation}
T(\nu,Q^2)=\frac{1}{\pi}\int_0^1d\rho\,\rho^2
\int_{\lambda_{\min}^2}^{\infty}d\lambda^2\int_{-1}^1dz\,
\frac{\psi(\rho,\lambda^2)}{Q^2+M^2\rho^2+\lambda^2-
2z\rho\sqrt{\nu^2+M^2Q^2}-i\epsilon}.
\label{4.4}
\end{equation}
In the complex $\nu^2$-plane this function has the cut along the
positive semiaxis of real~$\nu^2$. This cut starts at the point
$\nu^2_{\min}$ which is determined by
\begin{equation}
\sqrt{\nu^2_{\min}+M^2Q^2}\,=\min_{\{\lambda,\rho,z\}}
\biggl|\frac{Q^2+M^2\rho^2+\lambda^2}{2z\rho}\biggr|,
\label{4.5}
\end{equation}
that yields $\nu_{\min}={Q^2}/{2}$.

Thus, the DR at hand takes the form\footnote{This equation
generalizes the DR for the real Compton effect; cf., e.g., with
the results of Ref.~\cite{50}.}
\begin{equation}
T(\nu,Q^2)=\frac{1}{\pi}\int_{Q^4/4}^{\infty}\frac{d\nu_1^2}
{\nu_1^2-\nu^2-i\epsilon}W(\nu_1,Q^2).
\label{4.6}
\end{equation}

The DR for the $\nu$-odd structure functions can be derived from the
JL representation~\eqref{4.1} in a similar way.

In terms of the Bjorken variable $x=Q^2/2\nu$ DR~\eqref{4.6}
takes the form
\begin{equation}
T(\nu,Q^2)\equiv T(x,Q^2)=\frac{2}{\pi}
\int_0^1\frac{dx_1}{x_1}\frac{1}{1-(x_1/x)^2}W(\nu_1,Q^2).
\label{4.7}
\end{equation}
Thus, $T(x,Q^2)$ is an analytic function in the complex $x$-plane
with the cut along the segment $-1\leq x\leq1$. The DR~\eqref{4.7} is
suitable for establishing a link with the operator product
expansion~(OPE).

A natural scaling variable for the JL representation
reads~\cite{30},~\cite{51}
\begin{equation}
s=\frac{1}{2}\sqrt{\frac{Q^2(Q^2+4M^2)}{\nu^2+M^2Q^2}}\,.
\label{4.8}
\end{equation}
In terms of this variable one can represent the dispersion
integral~\eqref{4.7} in the following form:\footnote{It is worth noting
that in terms of another variable (e.g., the Nachtmann one
$\xi=2x/\bigl(1+\sqrt{1+x^24M^2/Q^2}\,\bigr)$) such a representation
does not exist.}
\begin{equation}
T(\nu,Q^2)=\frac{2}{\pi}\int_0^{1}\frac{ds_1}{s_1}
\frac{1}{1-(s_1/s)^2}W(\nu_1,Q^2).
\label{4.9}
\end{equation}

\subsection{Analytic moments of structure functions}
\label{ss4.3}

For the JL representation a natural scaling variable has the form of
Eq.~\eqref{4.8}. It can also be rewritten in terms of Bjorken
variable~$x$, namely,
\begin{equation}
s=x\sqrt{\frac{Q^2+4M^2}{Q^2+4M^2x^2}}\,.
\label{4.10}
\end{equation}
One can infer that for the physical processes the variable $s$
assumes the values in the range between~$0$ and~$1$.

The variable $s$ differs from both Bjorken and Nachtmann ones, which
are commonly employed in the analysis of the deep inelastic
scattering. However, it is this variable that leads to the moments
which possess appropriate analytic properties.

The $s$-moments of the structure functions can be defined as
\begin{equation}
{\mathcal M}_n(Q^2)=\frac{1}{{(1+4M^2/Q^2)}^{(n-1)/2}}
\int_0^{1}ds\,s^{n-2}W(\nu,Q^2).
\label{4.11}
\end{equation}
This function can also be rendered in the form of the
K\"allen--Lehmann representation
\begin{equation}
{\mathcal M}_n(Q^2)=(Q^2)^{n-1}\int_0^\infty
d\sigma\,\frac{m_n(\sigma)}{(\sigma+Q^2)^n},
\label{4.12}
\end{equation}
which elucidates its analytic properties. The support of the weight
function $m_n(\sigma)$ is located on the semiaxis $\sigma\geq0$. The
function $m_n(\sigma)$ can also be represented in terms of the
original distribution $\psi(\rho,\sigma)$ of the JL representation.

The analytic moments~\eqref{4.11} are related to the
$x$-moments~$M_n(Q^2)$ by~\cite{30}
\begin{equation}
{\mathcal M}_n(Q^2)=\frac{1}{\Gamma\bigl[(n+1)/2\bigr]}
\sum_{k=0}^{\infty}\frac{\Gamma\bigl[k+(n+1)/2\bigr]}{k!}
\biggl(-\frac{4M^2}{Q^2}\biggr)^kM_{n+2k}(Q^2).
\label{4.13}
\end{equation}

In the asymptotic high energy region $x$-, $s$-, and $\xi$-moments
coincide with each other, since the power corrections $1/(Q^2)^n$ can
be neglected. In the intermediate- and low-energy regions, where
higher twist contributions become considerable, one has to
distinguish between these moments.

The moments of the structure functions~\eqref{4.11}, being analytic
functions in the complex $Q^2$-plane with a cut, are convenient
objects for employment within the analytic approach to QCD. The
analyticity property is a consequence of the general principles of
the local~QFT.

\subsection{Relation with the operator product expansion}
\label{ss4.4}

The identity of the structures of DR in $x$ and $s$ variables allows
one to establish a relation of the analytic moments with the operator
product expansion. The latter is commonly employed in the study of
the $Q^2$-evolution of the structure function moments.

The $x$-moments of the structure function correspond to the case
when only the Lorentz-structures of the form $P_{\mu_1}\dots
P_{\mu_n}$ are retained in the matrix element
\begin{equation}
\langle P|\widehat{O}_{\mu_1\dots\mu_n}|P\rangle.
\label{4.14}
\end{equation}
In this case, the OPE for the Compton amplitude leads to the expansion
in powers of $(qP)/Q^2$, i.e., in the inverse powers of~$x$. The same
expansion in the inverse powers of~$x$ can also be performed in the
dispersion integral~\eqref{4.7}. The relevant coefficients are
determined by the $x$-moments. The comparison of these two expansions
provides one with the relation of the $x$-moments with the~OPE.

In a general case, the symmetric matrix element~\eqref{4.14} contains
the Lorentz-structures of the form
\begin{equation*}
\{P_{\mu_1}\dots P_{\mu_n}\},\quad
M^2g_{\mu_i\mu_j}\{P_{\mu_1}
\dots P_{\mu_{n-2}}\},\dots\,.
\end{equation*}
The $\xi$-moments correspond to the choice of the operator basis,
which involves the traceless tensors (i.e., such that the contraction
of the metric $g_{\mu_i\mu_j}$ with $\langle
P|\widehat{O}_{\mu_1\dots\mu_n}|P\rangle$ over any pair of indices
vanishes) as the expansion elements.

The dispersion relation~\eqref{4.9} allows one to expand the Compton
amplitude in inverse powers of~$s$. If the OPE basis is chosen in
such a way that an arbitrary contraction of the tensor $\langle
P|\widehat{O}_{\mu_1\dots\mu_n}|P\rangle$ with the nucleon momentum
$P_{\mu_i}$ vanishes, then the OPE leads to a power series for the
forward Compton scattering amplitude with the expansion parameter
$q_{\mu} q_{\nu}(P_{\mu} P_{\nu}-g_{\mu\nu}P^2) (q^2)^{-2}$ which
corresponds to expanding dispersion integral~\eqref{4.9} in the
inverse powers of~$s$. This establishes a relation of the analytic
moments of the structure functions with the OPE. It should be
stressed that the orthogonality requirement of the symmetric tensor
$\langle P|\widehat{O}_{\mu_1\dots\mu_n}|P\rangle$ to the nucleon
momentum $P_{\mu_i}$ unambiguously determines its Lorentz structure.

\subsection{Target mass effects}
\label{ss4.5}

The OPE method was applied to the description of the effects due
to the mass of the target in Ref.~\cite{52}. This approach leads to
the so-called $\xi$-scaling, i.e., the parton distributions become
functions of the $\xi$-variable. The resulting expressions for the
structure functions are inconsistent with the spectral condition.
This situation reminds the ghost pole problem, namely, that an
approximate solution contradicts the general principles of the
theory.

\begin{figure}[t]
\centering
\includegraphics{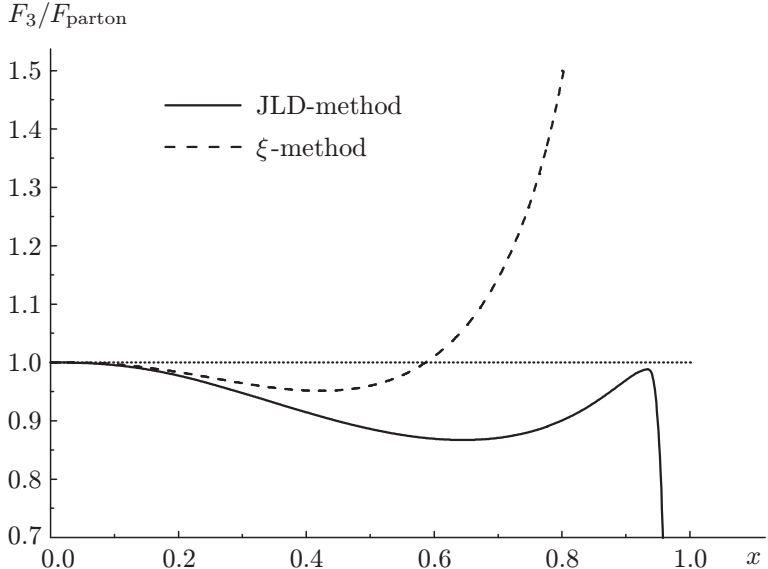}
\caption{Ratio of the structure function $F_3$, which retains the
massive corrections, to the parton distribution for~$M^2/Q^2=1/2$.
The solid curve corresponds to the incorporation of the effects due
to the target mass by making use of the JL representation. The dashed
curve corresponds to the $\xi$-scaling method.}
\label{fig8}
\end{figure}

The solution of the spurious singularity problem was based on the
K\"allen--Lehmann representation. One can avoid the contradiction of
the issue at hand with the spectral condition by accounting for the
target mass corrections in the framework of the JL representation.
This approach was implemented in Ref.~\cite{53}. Figure~\ref{fig8}
depicts the ratio of the structure function $F_3$, which retains the
mass corrections, to the parton distribution at $M^2/Q^2=1/2$. The
solid curve corresponds to the incorporation of the target mass
effects by making use of the JL representation, whereas the dashed
one corresponds to the $\xi$-scaling method. One can infer from
Fig.~\ref{fig8} that the structure function obtained within the
$\xi$-scaling method considerably deviates from that of the JL-method
at large values of~$x$. It is exactly the region ($x\sim1$) where the
$\xi$-method comes into contradiction with the spectral condition.
Since the accuracy of the experimental data is improving and such
subtle effects as higher twists become important, in the theoretical
studies one should originate in methods which are consistent with the
general principles of~QFT.

\section{Conclusion}
\label{sec5}

Renorminvariant PT is a basic tool for investigation of the QFT
models as well as for practical calculations of the elementary
particle interaction processes. The achievements of PT in QED and in
the electroweak theory are well known. The perturbative component of
QCD is substantial in the study of practically all the hadronic
processes. In the low-energy region, where the nonperturbative
effects become essential, the logarithmic contributions calculated
within PT are usually supplemented with the power corrections of the
nonperturbative origin. Thus, nonperturbative calculations involve PT
as a component. As a result, PT affects the determination (from
experimental data) of such nonperturbative quantities, as, for
example, vacuum expectation values of quark and gluon condensates.

The renormalized perturbative series contains large logarithms. Their
presence keeps the terms of the series large at high energies, even
when the expansion parameter (i.e., the coupling~$\alpha_{\mu}$) is
small. Nonetheless, since a perturbative expansion is not the final
result of the theory, its certain modification is admissible.

The RG method accumulates large logs within a new expansion
parameter, namely, the invariant charge $\bar\alpha_{\sa}(Q^2)$. This
results in perturbative expansions in powers of
$\bar\alpha_{\sa}(Q^2)$. The QCD invariant charge
$\bar\alpha_{\sa}(Q^2)$ is small at high energies, that constitutes a
crucial feature of QCD, namely, the asymptotic freedom (i.e., the
quark-gluon interaction is suppressed at small distances). Thus, the
principle of renorminvariance, which is implemented in the RG
method, allows one to modify the perturbative series. The resulting
expansions possess reasonable behavior in the UV domain (one has to
keep in mind the asymptotic character of the perturbative series).
These expansions are suitable for practical applications.

However, at this stage the perturbative expansions entail the
problems due to the unphysical singularities of the invariant charge.
Such singularities contradict the general principles of the local
QFT. The expansion in powers of $\bar\alpha_{\sa}(Q^2)$ becomes
unapplicable at small energies. Besides, at intermediate energies,
which are important in QCD applications, the perturbative results
contain large ambiguity due to the scheme dependence.

The APT constitutes the next step in the modification of the
perturbative expansion which eliminates the afore-mentioned
difficulties. Here the principle of renorminvariance is supplemented
with the causality principle. In the case of the quark current
correlation function the latter principle is implemented in the
analyticity condition which corresponds to the spectral
K\"allen--Lehmann representation. For the inelastic lepton-hadron
scattering the general properties of structure functions are
expressed in the integral Jost--Lehmann representation.

In conclusion, we summarize the most important features of APT:

$\bullet$ The effective Euclidean $\alpha_{\Ea}(Q^2)$ and Minkowskian
$\alpha_{\Ma}(s)$ invariant charges are defined in a self-consistent
way; they are free of unphysical singularities; they involve no
additional parameters; the higher APT functions $\mathcal A_k$ and
$\mathfrak A_k$ possess similar properties;

$\bullet$ Euclidean invariant charge $\alpha_{\Ea}$ as a function of
$\alpha_{\sa}$ obeys an essential singularity
$\exp(-1/\beta_0\alpha_{\sa})$ at the origin;

$\bullet$ It also has the infrared stable point
$\alpha_{\Ea}(0)=\alpha_{\Ma}(0)=1/\beta_0$ which
is independent of the scale parameter~$\Lambda$;

$\bullet$ In the framework of the APT the functional invariant
expansions in powers of $\bar\alpha_{\sa}$ for observables are
replaced by the nonpower expansions over the sets $\bigl\{\mathcal
A_k(Q^2)\bigr\}$ and $\bigl\{\mathfrak A_k(s)\bigr\}$;

$\bullet$ The dependence of the APT results on both the multi-loop
(NNLO and N$^3$LO) corrections and the choice of the renormalization
scheme is drastically suppressed in comparison with that of the
perturbative results.

We conclude this paper with the utterance which was the epigraph of
one of our papers:

\begin{center}
{\sl ``Take care of Principles and the Principles will take care of
you''.}
\end{center}

We hope that this sentence is consonant with the creative work
of Anatoly Alexeevich Logunov.

\kern-2mm

\subsection*{Acknowledgements}
This work was partially supported by the RFBR grant No.~05-01-00992,
Leading scientific schools grant No.~NS-5362.2006.2, BelRFBR grant
No.~F06D-002, and the Collaboration program between the research
centers of Belarus and JINR.

\end{document}